\documentstyle[preprint,eqsecnum,aps]{revtex}

\begin{document}
\draft
\title{Evidence for Skyrmions and Single Spin Flips in the
Integer Quantized Hall Effect}
\author{A. Schmeller, J.P. Eisenstein, L.N. Pfeiffer and K.W. West}
\address{AT\&T Bell Laboratories, Murray Hill, NJ 07974}
\date{\today}
\maketitle
\begin{abstract}
We have employed tilted-field magneto-transport measurements of the
energy gap for the odd-integer quantized Hall states at Landau
level filling factors $\nu = 1$,3 and 5 to determine the spin of
thermally excited quasielectron-quasihole pairs.  At $\nu = 1$ our
data show that as many as 7 electron spin flips accompany such
excitations, while at $\nu =3$ and 5 apparently only a single spin
flips.  These results lend support to the recent suggestions that
"Skyrmionic" quasiparticles are the lowest-lying charged
excitations of the fully-polarized $\nu = 1$ quantum Hall fluid but
are not at the higher odd-integer fillings.
\end{abstract}
The integer and fractional quantized Hall effects (QHE) which occur
in two-dimensional electron systems (2DES) at high magnetic field
are customarily distinguished by the origin of the underlying
energy gaps in the two cases\cite{QHEbooks}.  On the one hand,
the magnetic field-induced resolution of the single-particle energy
spectrum
into a series of discrete, but highly degenerate, spin-split Landau
levels provides all the energy gaps necessary to explain the
integer QHE, with each such Hall plateau corresponding to the
complete filling of an integer number of these single-particle
levels. In contrast, for the fractional effect, which occurs only
at certain partial fillings of the Landau levels, the requisite
energy gap derives entirely from many-body effects.  This
distinction, while often useful, can be misleading, especially in
the case of the odd-integer QHE.  For these states, in which the
Fermi level resides in the spin-flip gap within the uppermost
Landau level, electron-electron interaction effects have long been
known\cite{nicholas,usher} to greatly enhance the energy gap above
the single-particle Zeeman energy.  In fact, it is now
believed that the odd-integer QHE states would survive even if the
Zeeman energy were removed entirely\cite{sondhi1}.  In this case, the
2DES would develop spontaneous ferromagnetic order (at zero
temperature) solely because of interaction
effects.  But perhaps even more remarkable than this, are the
recent predictions concerning the nature of the elementary
excitations of these ferromagnetic states\cite{sondhi1,fertig}.
Provided that the Zeeman energy is sufficiently small, the
lowest-lying charged excitation is not simply a single flipped spin
but is instead a large, smooth distortion of the spin field in
which many spins are flipped.  While such excitations, whose charge
is $\pm e$, obviously cost more Zeeman energy than a single spin
flip, the near-parallelism of neighboring spins saves on exchange
energy. The total spin, and hence the spatial extent, of these
objects is determined by the competition between these two
energies.  Evidence for these unusual excitations (known as
"Skyrmions" in the limit of zero Zeeman energy) has recently been
uncovered in NMR Knight-shift studies of the 2DES ground state spin
polarization\cite{barrett}.  In this Letter
we report on transport studies which directly probe the {\it charged}
excitations of odd-integer quantized Hall states.  Our findings
strongly suggest that while large-spin Skyrmionic quasiparticles
dominate the $\nu =1$ integer QHE (where only the lower spin
branch of the lowest Landau level is occupied), they are not
relevant to the higher odd-integer states at $\nu =3$ and 5.

In this experiment we determine the energy gap $\Delta$ for
creating a widely-separated quasielectron-quasihole pair in a given
quantized Hall state by measuring the temperature dependence of the
longitudinal resistance in the thermally activated regime where $R_{xx}
=R_0 exp(-\Delta /2T)$.  We then assume that the spin $s$ of
the quasiparticle pair can be extracted from the change in $\Delta$
produced by tilting the total magnetic field $B_{tot}$ away from
normal to the 2D plane (keeping, however, the perpendicular field
$B_\perp$, and thus the Landau level filling factor $\nu$,
fixed). The basis for
this assumption is that for an ideal, infinitely thin 2DES, an
in-plane magnetic field $B_\parallel$ couples to the system
{\it only} through the Zeeman energy, while the perpendicular field
 controls the orbital dynamics\cite{fangstiles}.  This implies
that the Zeeman contribution to the energy gap $\Delta$ for
creating a quasiparticle pair with spin $s$ (in units of $\hbar$)
out of a polarized ground state is merely additive:
\begin{equation}
\Delta = \Delta_{0,s} (B_\perp ) + s |g| \mu_B B_{tot}
\end{equation}
The first term, $\Delta_{0,s}(B_\perp )$, is the contribution to
the gap arising from all non-Zeeman sources (e.g. many-body
effects) and, in this model, depends only upon the perpendicular
magnetic field $B_{\perp}$.  Hence, the derivative $\partial \Delta
/ \partial B_{tot}$ (evaluated at constant $B_\perp$) is just
$s|g|\mu_B$.  Since the $g$-factor and Bohr magneton $\mu_B$ are
known ($|g| \approx 0.44$ in GaAs\cite{dobers} and $\mu_B =0.672K/T$),
measuring $\partial \Delta / \partial B_{tot}$ determines the spin
$s$.  For example, in the traditional (i.e. pre-Skyrmion) view of
the fully spin-polarized $\nu =1$ QHE state, the lowest-lying
charged excitation is assumed to be a single flipped spin,
i.e. $s=1$.  In this case $\Delta_{0,1}$ is the Coulomb exchange
energy\cite{fczhang} $E_{ex} = \sqrt{\frac{\pi}{2}}e^2 /\epsilon\ell_0$
(where $\epsilon \approx 13$ is the dielectric constant of GaAs and
$\ell_0=\sqrt{\hbar/eB_{\perp}}$ is the magnetic length).  Although
the exchange term dominates the net gap in typical GaAs/AlGaAs
systems, the Zeeman contribution should nevertheless be detectable
via the derivative $\partial \Delta / \partial B_{tot} = +|g|
\mu_B$ (=0.3K/T in GaAs).

The four samples used in this study are modulation-doped
GaAs/AlGaAs heterostructures grown by molecular beam epitaxy.  Two
are conventional single heterointerfaces, and two are GaAs single
quantum wells, with widths of 200\AA$\:$ and 140\AA$\:$,
respectively.
As grown, these samples, labeled SI1, SI2, QW1, and QW2, have 2DES
densities of N=0.6, 1.3, 2.1, and ${\rm 1.4 \times 10^{11} cm^{-2}}$,
and low temperature mobilities of $\mu =$3.4, 2.8, 2.0, and ${\rm
0.38 \times 10^6 cm^2 /Vs}$.  In addition, for samples QW1 and QW2,
these parameters could be altered significantly by applying voltages
to metal gate electrodes placed on the sample's top and/or bottom
surface.  Each sample was a roughly 5x5mm square with eight
diffused In contacts around the outer edge.  Conventional
magneto-transport measurements were performed down to 0.5K using
100nA, 5Hz excitation.  Tilting of the samples with respect to the
applied magnetic field was performed {\it in-situ} at low temperature.

Figure 1 shows typical temperature dependences of the resistivity
minimum at $B_\perp =2.3T$ of the $\nu =1$ QHE state in sample
SI1.  Data obtained with the magnetic field perpendicular to the 2D
plane ($\theta =0$) and tilted out to $\theta =56^\circ$ are shown.
The dashed lines are least-squares fits to the linear portion of
the data; from the slopes of these lines we find energy gaps of
$\Delta =19$ and 23K for $\theta =0$ and $56^\circ$, respectively.
These gap values are much larger than the Zeeman energy in
GaAs ($|g| \mu_B B_{tot}=0.7K$ at $B_{tot}=2.3T$) and clearly
demonstrate the well-known\cite{nicholas,usher} dominance of
many-body effects at $\nu =1$.  Figure 2 shows the overall tilted field
dependence of the $\nu =1$ energy gap observed in samples SI1, QW1
and QW2. (For sample QW1 the two data sets shown were obtained
using different gating configurations \cite{gate}.) The leftmost point in each
data set corresponds to $\theta =0$ and $B_{tot}=B_\perp$.  The
energy gap $\Delta$ is given in units of $e^2/\epsilon \ell_0$ and
the total magnetic field is represented by the dimensionless Zeeman
energy $\widetilde{g}\equiv |g| \mu_B B_{tot}/(e^2/\epsilon
\ell_0)$. (The Coulomb energy $e^2/\epsilon \ell_0$ depends only
upon $B_{\perp}$ and is thus constant in a tilt experiment.)

As Fig. 2 shows, the $\nu =1$ energy gaps initially rise quickly as
the magnetic field is tilted.  The initial slope $\partial \Delta
/\partial B_{tot}$ is roughly 2K/T and is the same in all the
samples.  This slope is some 7 times larger than what Eq. 1
predicts for excitation of quasiparticle pairs involving a single
spin flip.  Assuming that Eq. 1 gives an adequate basis for
interpreting these data, the large slope suggests that unusual
large-spin ($s \approx 7$) charged objects are being thermally
excited.  On the other hand, Nicholas, {\it et al.}\cite{nicholas},
who first noticed the large $\partial \Delta /\partial B_{tot}$ at
$\nu =1$, attributed it to a breakdown within the assumptions
underlying Eq. 1.  But before arguing that our results do in fact
imply the existence of large-spin charged excitations at $\nu =1$, we
turn to our experimental results for the higher-order spin-flip QHE
states at $\nu =3$ and 5.

Figure 3 displays the tilted field results for $\nu =3$ and 5
obtained using samples SI2, QW1 and QW2.  For these samples the
2DES density (adjusted by gating, if necessary) produced these
higher filling factors at about the same perpendicular magnetic
field as employed earlier at $\nu =1$.  This assures that the
various energy scales (Zeeman, Coulomb, and cyclotron) are all of
the same magnitude as they were for $\nu =1$.  Again, the figure
plots the normalized energy gap $\Delta/(e^2/\epsilon \ell_0)$
versus dimensionless Zeeman energy $|g| \mu_B B_{tot}/(e^2/\epsilon
\ell_0)$.  While the observed gaps at these filling factors are
somewhat smaller than that found at $\nu =1$, they still exceed the
Zeeman energy $|g| \mu_B B_{tot}$ by about an order of magnitude.
Thus, interaction effects dominate these integer QHE states as
well.  On the other hand, instead of a large initial slope (the
dashed line corresponds to the slope seen at $\nu =1$), the data in
Fig. 3 exhibit rather little variation with tilt.  The dotted lines
have slopes appropriate to single spin flips ($s=1$) and, while
$\nu =3$ in sample QW2 and $\nu =5$ in sample QW1 are consistent
with this, the two other data sets show even weaker dependences.
Thus, we observe a qualitative difference between the
charged excitations of the $\nu =3$ and 5 QHE states and those at
$\nu =1$.

We believe that the data shown in Figs. 2 and 3 strongly support
the recent theoretical predictions about Skyrmionic excitations in
the quantized Hall effect.  Nevertheless, before comparing our data
to these predictions, we first discuss two effects not present in
an ideal, infinitely thin 2DES, and show that they cannot be held
responsible for our results.  We consider first the non-Zeeman
effects of large in-plane magnetic fields. Owing to the finite
thickness of real 2D electron systems, an in-plane magnetic field
couples not only to the Zeeman energy, but also to the
perpendicular dynamics.  This coupling involves mixing between the
subbands of the heterostructure confinement potential. Although the
effect of these mixings on QHE gaps is not well
understood\cite{mixing}, the controlling parameter is the thickness
of the 2D sheet or,
equivalently, the energy splitting between the lowest and
first-excited confinement subbands.  For sample SI1, a conventional
single heterointerface, self-consistent solution of the
Schroedinger and Poisson equations\cite{LDA} yields an estimated subband
splitting of $E_{01} \approx 7{\rm meV}$ and an rms thickness for
the ground subband wavefunction of $\sigma_z \approx {\rm 76\AA}$.
In order to have a significantly thinner 2DES, with its
concomitantly larger subband splittings, we chose to study
quantum well samples.  For the ${\rm 200\AA}$ quantum well
sample QW1 $E_{01} \approx 31{\rm meV}$ and $\sigma_z \approx {\rm
42 \AA}$ while in the ${\rm 140\AA}$ sample QW2 these numbers are
57meV and ${\rm 31\AA}$, respectively.  But, as Fig. 2 clearly
demonstrates, for these much thinner 2DES samples the slopes
$\partial \Delta /\partial B_{tot}$ observed at $\nu =1$ are nearly
the same as that observed in sample SI1.  The same conclusion
applies to the $\nu =3$ and 5 data in Fig. 3, although the
anomalously small slope for $\nu =3$ in sample SI2 may in fact be a
residual finite-thickness effect.  In our opinion, the thickness
independence of our results at $\nu =1$,3, and 5 strongly discounts
subband mixing effects of the in-plane magnetic field.

Another important question concerns the role of disorder in the
2DES.  Indeed, the samples used here differ significantly in their
zero field mobility $\mu$: for sample SI1 $\mu ={\rm 3.4 \times 10^6
cm^2 /Vs}$ while sample QW2, when gated as in Fig. 2, has $\mu
={\rm 1.6 \times 10^5 cm^2 /Vs}$.  Although the mobility is not
necessarily the best measure of the disorder relevant to the QHE,
the observed $\nu =1$ gap magnitude is systematically smaller in
the samples with lower mobility.  But, as already noted, the tilted
field behavior displayed in Figs. 2 and 3 are the same from one
sample to the next.  This is strong evidence that disorder is not
playing a qualitatively important role.  In particular, it argues
against the suggestion\cite{nicholas} that the large slope at $\nu =1$
is due to incomplete spin polarization of the $\nu =1$ ground state
resulting from the overlap of disorder-broadened spin branches of
the lowest Landau level.  (Indeed, were such a mechanism
operative, one would expect large slopes at $\nu =3$ and 5 as well
as $\nu =1$.)

While neither finite thickness nor disorder effects appear to
explain the large $\partial \Delta /\partial B_{tot}$ observed at
$\nu =1$, our results are in excellent qualitative agreement with
recent theory\cite{sondhi1,fertig} which predicts that the
lowest-lying charged excitations at $\nu =1$ are large-spin
Skyrmionic quasiparticles.  The spin $s$ of a thermally excited
Skyrmion/anti-Skyrmion pair depends upon the ratio $\widetilde{g}$
of Zeeman to Coulomb energies.  For sufficiently large
$\widetilde{g}$, $s=1$ and "Skyrmions" are identical to single spin
flips.  In this limit the predicted energy gap (in units of
$e^2/\epsilon \ell_0$) is
$\widetilde{\Delta}=\sqrt{\frac{\pi}{2}}+\widetilde{g}$.  In the
opposite limit, $\widetilde{g} \to 0$, the Skyrmion spin and size
diverge while the energy gap approaches
$\widetilde{\Delta}={\frac{1}{2}}\sqrt{\frac{\pi}{2}}$,
i.e. precisely one half the energy required to flip a single
spin\cite{sondhi1}.
The inset to Fig. 2 displays one calculation\cite{fertig,macd} of the
$\nu =1$ energy gap as a function of $\widetilde{g}$ for an ideal,
infinitely thin, 2DES.  This Hartree-Fock calculation, which
asymptotically approaches the single spin flip at large
$\widetilde{g}$, does not adhere to the expectation\cite{sondhi2} that
the spin of the Skyrmion/anti-Skyrmion pair is always an odd integer
and that, as a result, the energy gap is actually a continuous
sequence of straight lines segments, each with slope
$\partial\widetilde{\Delta}/\partial\widetilde{g}=s={\rm
1,3,5,...}$.  Nevertheless, it is apparent that the general shape
of the theoretical curve in the inset is in good qualitative
agreement with our experimental results.  Furthermore, the spin
size we infer, $s\approx 7$ at $\widetilde{g} \approx 0.01$ is
close to the theoretical estimate\cite{fertig,macd} of $s\approx
9$.  On the other hand, the magnitude of the measured gap itself is
only about 25\% of the theoretical value.  There are, however, several
possible sources of energy gap suppression, including disorder,
finite thickness effects, and Landau level mixing.  While the
effect of disorder is probably small in sample SI1, which has
a mobility in excess of $3 \times 10^6 {\rm cm^2 /Vs}$,
estimates\cite{pseudo} suggest that the thickness-induced softening
of the Coulomb interaction reduces the $\nu =1$ gap by roughly 30\%.
Landau level mixing, however, may be the most important effect
since for our samples, in which $\nu =1$ occurs at $B_{\perp}
\approx 2T$, the Coulomb energy $e^2/\epsilon \ell_0$ actually
$exceeds$ the cyclotron gap.  Indeed, recent variational
calculations\cite{louie} suggest that Landau level mixing can
reduce the $\nu =1$ energy gap very substantially ($\sim 50\%$) at
low magnetic fields.  In view of these considerations, we do not
believe that the quantitative disagreement between Hartree-Fock
theory and our experiment invalidates our fundamental conclusion
that large-spin charged excitations dominate the $\nu =1$ QHE gap.

Remarkably, recent theoretical work by Wu and Sondhi\cite{wu}
predicts that within the integer QHE, Skyrmions are the lowest-lying
charged excitations ${\it only}$ for the case $\nu =1$.  Even
without the Zeeman energy, conventional single spin flips are
predicted to be lower in energy than Skyrmions for all $\nu \ge
3$.  In spite of the great similarity between the higher
odd-integer filling factors and $\nu =1$, Skyrmions are apparently
destabilized by the subtly different electron-electron interactions
in the higher Landau levels.  Although Wu and Sondhi\cite{wu}
ignore the possibly important effects of Landau level mixing, our
observation of only very small slopes $\partial \Delta /\partial
B_{tot}$ at $\nu =3$ and 5 appears to qualitatively verify their
prediction.

In conclusion, we have used tilted field studies of the energy gap
for the $\nu =1$, 3 and 5 integer quantized Hall states to estimate
the spin of thermally excited quasielectron-quasihole pairs.  At
$\nu =1$ our results reveal unusual charged excitations in which
typically 7 spins are reversed at $B_\perp =2.3T$.  In contrast, at
$\nu =3$ and 5 our findings are consistent with ordinary single
spin flip excitations. Both of these results are in excellent
qualitative agreement with recent theory on Skyrmionic excitations
in the quantized Hall effect.

It is a pleasure to thank Luis Brey, Song He, Allan
MacDonald, and Shivaji Sondhi for useful discussions.  We also
thank Song He for performing informative small system exact
diagonalization calculations at our request.


\begin{figure}
\caption{Arrhenius plots of the longitudinal resistance $R_{xx}$ at filling
factor
$\nu =1$ ($B_{\bot} =2.3$ T) for sample SI1. The data sets are recorded for
tilt angles
$\Theta =0^{o}$ and $56^{o}$. The experimental geometry is shown in the lower
left inset.
The upper right inset displays traces of $R_{xx}$ at $\theta =0$ versus
magnetic
field around $\nu =1$.}
\label{fig1}
\end{figure}

\begin{figure}
\caption{Results of tilted field experiments on the $\nu =1$ QHE.
The energy gaps $\Delta$ at fixed $B_{\bot}$ are plotted versus the
Zeeman energy $g\mu_{B} B_{tot}$, both in units of $e^{2}/\epsilon \ell_{0}$.
Each data set starts with $\theta =0$ and $B_{tot} = B_{\bot}$ at the lower
left.
On the quantum well samples we use gate electrodes to tune the electron
densities \protect\cite{gate}.
{}From top to bottom the samples had
electron densities 0.6, 1.0, 0.6, and $1.0\times 10^{11}$cm$^{-2}$ and
 mobilities 3.4, 0.52, 0.18, and $0.16\times 10^{6}$cm$^{2}$/Vs, respectively.
For comparison we include lines with
$\partial \Delta / \partial (g\mu_{B} B_{tot})=s=7$ (dashed) and 1 (dotted).
The inset shows a Hartree-Fock result of Skyrmion theory (full line)
\protect\cite{fertig,macd}. }
\label{fig2}
\end{figure}

\begin{figure}
\caption{Energy gaps versus Zeeman energy as in Fig. 2, but now at filling
fractions
$\nu =3$ and 5 (as indicated). Again, we include lines with $s=1$ (dotted) and
7 (dashed)
for comparison. The sample parameters for the data sets from top to bottom
 were:
electron densities 1.3, 1.4, 2.1, and $2.1\times 10^{11}$cm$^{-2}$;
 mobilities 2.8, 0.38, 2.0, and $0.48\times 10^{6}$cm$^{2}$/Vs.}
\label{fig3}
\end{figure}

\begin{references}
\bibitem{gate}
Sample QW1 was fitted with both front and back gates.  It was thus
possible to obtain different mobilities at the same density by
forcing the 2DES wavefunction toward the upper (cleaner) or lower
(dirtier) interface of the quantum well.
\end{references}
\end{document}